\documentclass[CEJP,PDF]{cej} 
\usepackage{layout}
\usepackage{amsmath}
\usepackage{textcomp}
\usepackage{hyperref}

\title{Low Temperature Measurements by Infrared Spectroscopy in CoFe$_2$O$_4$ Ceramic}

\articletype{Research Article}

\author{Renata Bujakiewicz-Koro\'{n}ska\inst{1}\email{rbk@up.krakow.pl},
        	\L ukasz Hetma\'{n}czyk\inst{2,3},
        	Barbara Garbarz-Glos\inst{1},
			Andrzej Budziak\inst{4},
			Anna Kalvane\inst{5},
			Karlis Bormanis\inst{5},
			Kacper Dru\.{z}bicki\inst{2,3}}

\institute{
    \inst{1}Pedagogical University, Institute of Physics, Podchorazych 2, 30-084 Krakow, Poland
    \inst{2}Jagiellonian University, Faculty of Chemistry, Ingardena 3, 30-060 Krakow, Poland
    \inst{3}Frank Laboratory of Neutron Physics, Joint Institute for Nuclear Research, 141980 Dubna, Russia
    \inst{4}The H. Niewodniczanski Institute of Nuclear Physics PAN, Radzikowskiego 152, 31-342 Krakow, Poland
    \inst{5}Institute of Solid State Physics, University of Latvia, Kengeraga 8, LV-1063, Riga, Latvia }

\abstract{In this paper results of new far-infrared and middle-infrared measurements (wavenumber range of 4000~cm$^{-1}$~-~100~cm$^{-1}$) in the range of the temperature from 300 K to 8 K of the CoFe$_2$O$_4$  ceramic are presented. The bands positions and their shapes are the same in the wide temperature range. The quality of the sample was investigated by X-ray, EDS and EPMA studies.  The CoFe$_2$O$_4$ reveals the cubic structure ($Fd-3m$) in the temperature range from 85~K to 360~K without any traces of distortion. On the current level of knowledge the polycrystalline CoFe$_2$O$_4$ does not exhibit phase transition in the temperature range from 8~K to 300~K.}


\begin{document}
\maketitle


\section{Introduction}

Cobalt spinel ferrite CoFe$_2$O$_4$ (CFO) is interesting due to its structural, magnetic, and electrical properties. It has Curie temperature around 793 K, strong magnetic anisotropy, high moderate saturation magnetization and coercivity, and high mechanical hardness. These properties with their physical and chemical stability make it suitable for magnetic recording applications \cite{journal-1, journal-2,journal-3,journal-4,journal-5,journal-6}.

The cobalt ferrite is extensively investigated. Raman and IR spectra for CFO were obtained in the range (800~-~200)~cm$^{-1}$ by Yu \textit{et al.} \cite{journal-7} at the temperature range (300~-~870) K and in the range (900~-~100) cm$^{-1}$ at room temperature by Wang \textit{et al.} \cite{journal-8}. The aim of this work is to study the low temperature dependence of FIR and MIR spectra in the range (4000~-~100)~cm$^{-1}$.

\section{Experiment and Results}
\subsection{Sample preparation}
The CFO samples were made by a conventional method. Powders of CoFe$_2$O$_4$ solid solutions were obtained by solid phase synthesis from  Co$_2$O$_3$-99,5~\% and Fe$_2$O$_3$-70 \% oxides. The starting materials were weighed according to the chemical formula, homogenized and milled in an agate ball-mortar in ethanol for 24 h, dried and calcined for 1 h at the temperature    1000~K. The calcined powders were next reground, pressed under the pressure 15 MPa, and sintered for 4 hours at 1100~K.

\subsection{X-ray investigations}
\begin{figure}
\includegraphics[width=0.7\textwidth]{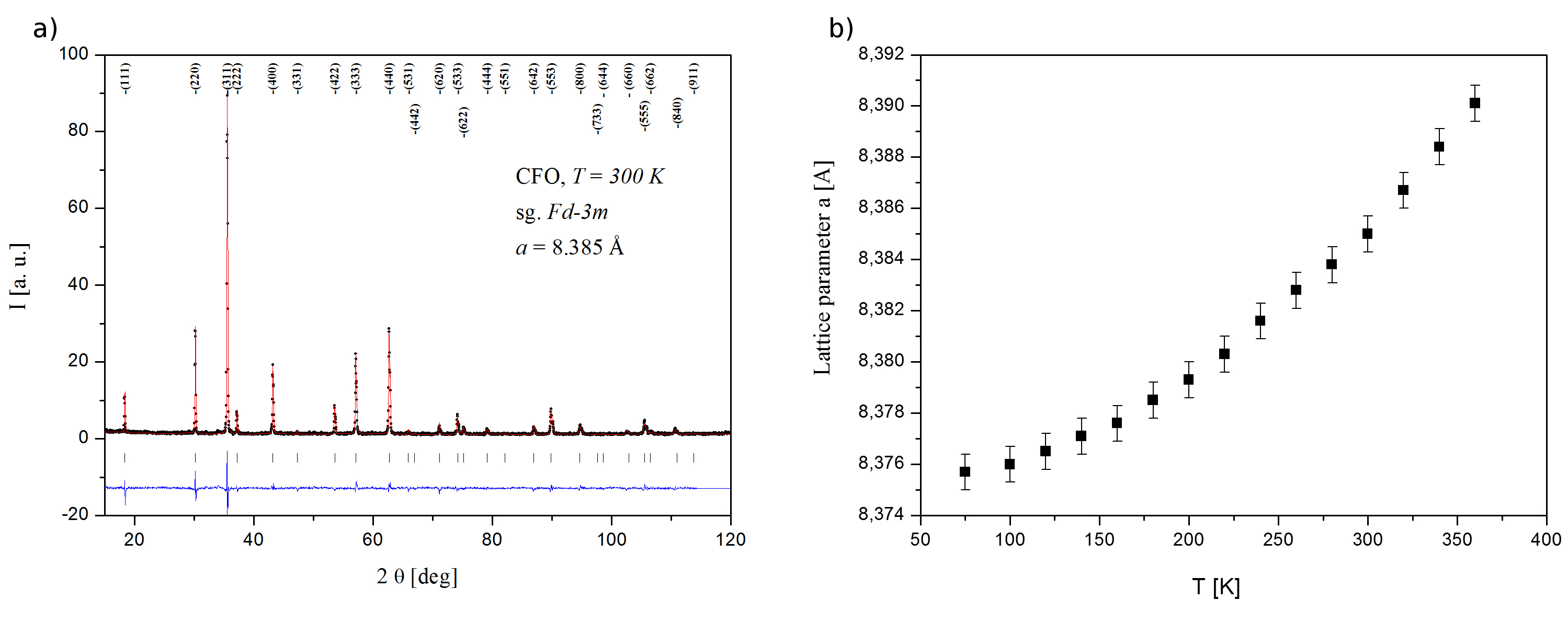}
\caption{a)~X-ray diffraction pattern of the CFO ceramic at RT. Upper tick marks indicate the position of the Bragg reflections for the cubic phase (\textit{Fd-3m}); only the strongest ones are described.
b)~The unit cell parameter $a$ of the cubic phase  (\textit{Fd-3m}) for the CFO as a function of temperature.
\label{fig1ab}}
\end{figure}

X-ray studies have been carried out using an X'Pert PRO (PANalytical) diffractometer with the CuK$_\alpha$ radiation and a graphite monochromator. Temperature of the CFO sample was stabilized (accuracy $\pm$1 K) by means of the low temperature chamber TTK 450. The measurements were performed during heating. After each heating stage the sample was held about 10min to obtain temperature equilibrium. A profile-fitting program FULLPROF~\cite{journal-9} based on the Rietveld method was used to analyze and fit the spectra. An exemplary diagram of X-ray measurement made at RT in Fig.\ref{fig1ab}a is presented  and the obtained lattice parameter   $a$ = 8.385~\AA~is in good agreement with data in \cite{journal-10}. The CFO sample reveals the cubic structure ($Fd-3m$) in the whole temperature range  (85~-~360) K without any traces of distortion. The lattice parameter increases from 8.3757~\AA~at 85 K up to 8.3901~\AA~at 360 K (Fig.\ref{fig1ab}b), which corresponds to the change of the unit cell volume $\sim$ 0.5 \%. Cations of Co and Fe occupy $8a$ and $16d$ special Wyckoff positions at (1/2,1/2, 1/2)  (1/8, 1/8, 1/8), respectively. Oxygen ions occupy the $32e$ positions at ($x, x, x$). The obtained value $x$ = 0.2555(8) at RT is in very good agreement with data from~\cite{journal-20}. Typical reliability factors obtained from refinements of the XRD patterns were: R$_p$ $\sim$ 12, R$_{wp}$ $\sim$ 15, R$_{wp}$ $\sim$ 6 and Chi$^2$ $\sim$ 5.5.

\subsection{Morphological and chemical investigations }
\begin{figure}
\includegraphics[width=1.0\textwidth]{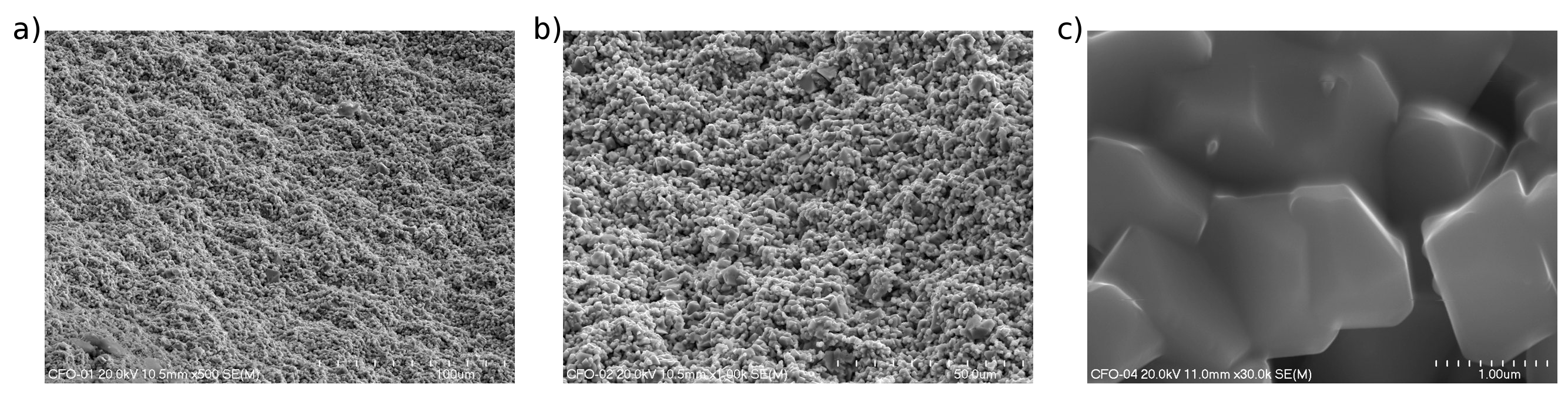}
\caption{SEM micrographs of the fractured surface of the CFO sample. Magnif.: a) 500$\times$, b)1000$\times$, c) 30000$\times$.\label{fig2}}
\end{figure}

The ceramic microstructure was investigated by means of an electron scanning microscope with microanalyses system Noran-Vantage. SEM micrographs of the fractured surface of the CFO sample with magnifications 500$\times$, 1000$\times$, and 30 000$\times$ are presented in Fig.~\ref{fig2}. The microphotographs of fractures of CFO showed that the sample was perfectly sintered and dense (95 \% theoretical). The fracture had a fragile nature and in grains the crystalline structures were observed. The average values of grain size (measured by the linear intercept method) was 300~nm. The homogeneity of element distribution in the samples was confirmed by the EPMA method using an X-ray microprobe. 

\subsection{Elastic properties}

Material constants: the Young's modulus $E$, the shear modulus $G$ and the Poisson ratio $\nu$  were determined based on the investigations of elastic properties of the CFO polycrystalline samples. An ultrasonic method was applied. The measurements were carried out with the INCO -VERITAS Ultrasonic Measuring Set UZP-1. Two kinds of transducers were used. The 10 MHz frequency transducers connected to a sample by oil were used for longitudinal waves while the 2 MHz frequency transducers stuck on a sample with Canada balsam was applied for transverse waves. Material constants values were calculated from a longitudinal and transverse ultrasonic waves propagation velocity and the apparent density of samples using the following formulas \cite{journal-11,journal-12}:
$$E = V_L^2\rho(1+\nu)(1-2\nu)/(1-\nu),$$
$$G = V_T^2 \rho, $$
$$\nu = (V_L^2 - 2V_T^2)/(2V_L^2 -2V_T^2),\label{w.1}$$
where: $\rho$ - density, $V_L$ - velocity of longitudinal wave, $V_T$ - velocity of transverse wave. $V_L$ achieved value 6189.4~m/s, and the next one $V_T$ was 3265.9~m/s. The measured parameters in a direction in the plane of the CFO ceramic are given in Table~\ref{tab1}. The CFO sample has high density and small porosity. Therefore material constants of  CFO are a bit different for these ones in \cite{journal-10}. These results confirmed high mechanical hardness of CFO.

\begin{table}
\caption{ Values of the elastic parameters of CFO.\label{tab1}}
\begin{tabular}{cccc}
\hline
$\rho$ [kg/m$^3$] & $E$ [GPa] & $G$ [GPa] &$\nu$ \\
\hline \hline
6260 $\pm$ 100 &  174.5 $\pm$ 16.6 &   66.8 $\pm$ 0.2 & 0.307 $\pm$ 0.014\\
\hline
\end{tabular}
\end{table}
\subsection{FT-FIR and FT-MIR studies}

The far (FIR) and middle (MIR) infrared absorption measurement were performed using  the Bruker 70v vacuum Fourier Transform spectrometer. The transmission spectra were made with a resolution of  2 cm$^{-1}$ and with 32 scans per each spectrum. The FIR spectra (600-80)~cm$^{-1}$ were collected for sample suspended in apiezon N grease placed on polyethylene disc. The MIR spectra (4000~-~400) cm$^{-1}$ were carried out for sample mixed with dry KBr and pressed into the form of pellet. 

The sample was loaded in cryostat DE-202A at room temperature and measurements were performed in the cooling process to ca. 8 K with the cooling rate 3 K/min. The desired temperature was measured with accuracy of 0.1 K and stabilized for ca. 3 minutes before the data were taken. The LakeShore 331S temperature controller equipped with diode sensor to control the temperature was used. The PE windows were used in cryostat in case of FIR measurements and KRS window in case of MIR measurements. 

\begin{figure}
\includegraphics[width=1.0\textwidth]{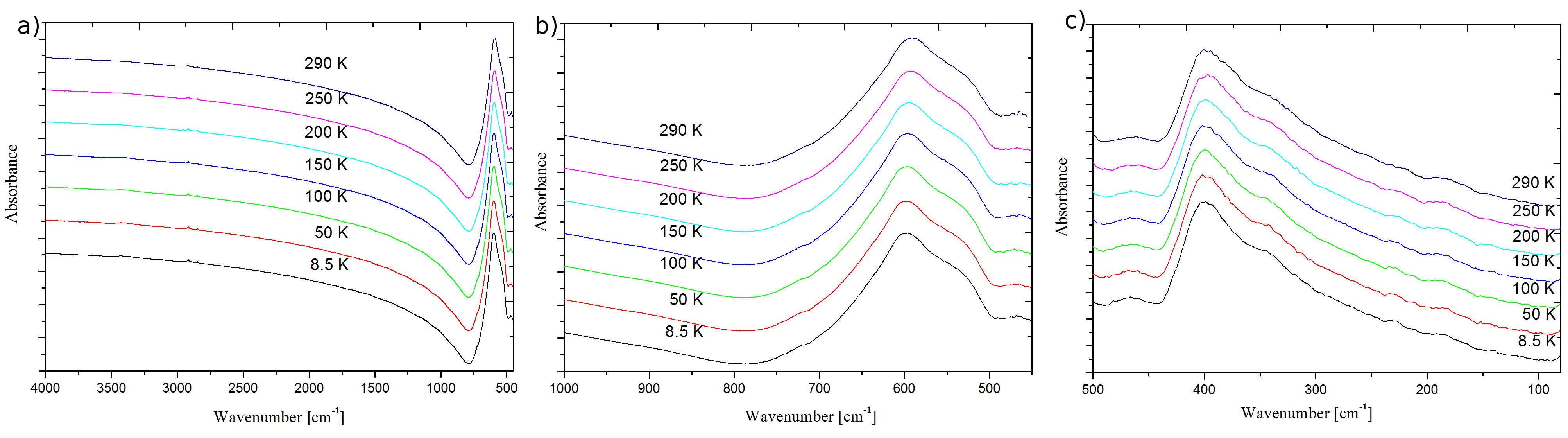}
\caption{Selected spectra for CFO in the region of  a) FT-MIR (4000-450) cm$^{-1}$, b) FT-MIR (1000-450) cm$^{-1}$, c) FT-FIR (500-80) cm$^{-1}$.\label{fig3}}
\end{figure}

Fig.~\ref{fig3}a presents selected MIR spectra in the region of (4000~-~450) cm$^{-1}$. One intense maximum is found at ca. 590 cm$^{-1}$. At lower wavenumbers it is clearly visible shoulder of the main band. None changes is visible during cooling. Peak position as well as band shape (full width at half maximum) does not change as a function of the temperature. Raman spectra were recorded and interpreted by Wang and Ren \cite{journal-8}, T. Yu \textit{et. al.}~\cite{journal-13} and P. Chandramohan \textit{et al.} \cite{journal-14}. Fig.~\ref{fig3}b shows selected MIR spectra in the region of (1000~-~450)~cm$^{-1}$. The presence of shoulder band is clearly visible. Fig.~\ref{fig3}c shows selected FIR spectra of the CFO compound. Very weak and broad absorption band is observed at 467 cm$^{-1}$. Next intense and broad band at ca. 400 cm$^{-1}$ is visible and shoulder band at ca. 345 cm$^{-1}$. Very weak and broad bands at ca. 230 cm$^{-1}$ and ca 180 cm$^{-1}$ are also visible. However, as can be noticed from this figure, one cannot observe any changes. The band positions and its shape are the same in the wide temperature range.

The bands at 590 cm$^{-1}$ and 467 cm$^{-1}$ are Raman active modes too. They are consistent with  575.5 cm$^{-1}$~\cite{journal-14} and 470 cm$^{-1}$~\cite{journal-8,journal-13,journal-14}. There are found four infrared active modes  F$_{1u}$ (400 cm$^{-1}$, 345 cm$^{-1}$, 230 cm$^{-1}$, 180~cm$^{-1}$). These low frequency modes can be assigned to vibrations of the tetrahedral sublattice, whereas the higher energy phonon mode at 590~cm$^{-1}$ probably is connected with vibrations of the octahedral sublattice ~\cite{journal-8}.

\section{Calculations}
\subsection{\textit{Ab initio} calculations}

\begin{figure}
\includegraphics[width=0.9\textwidth]{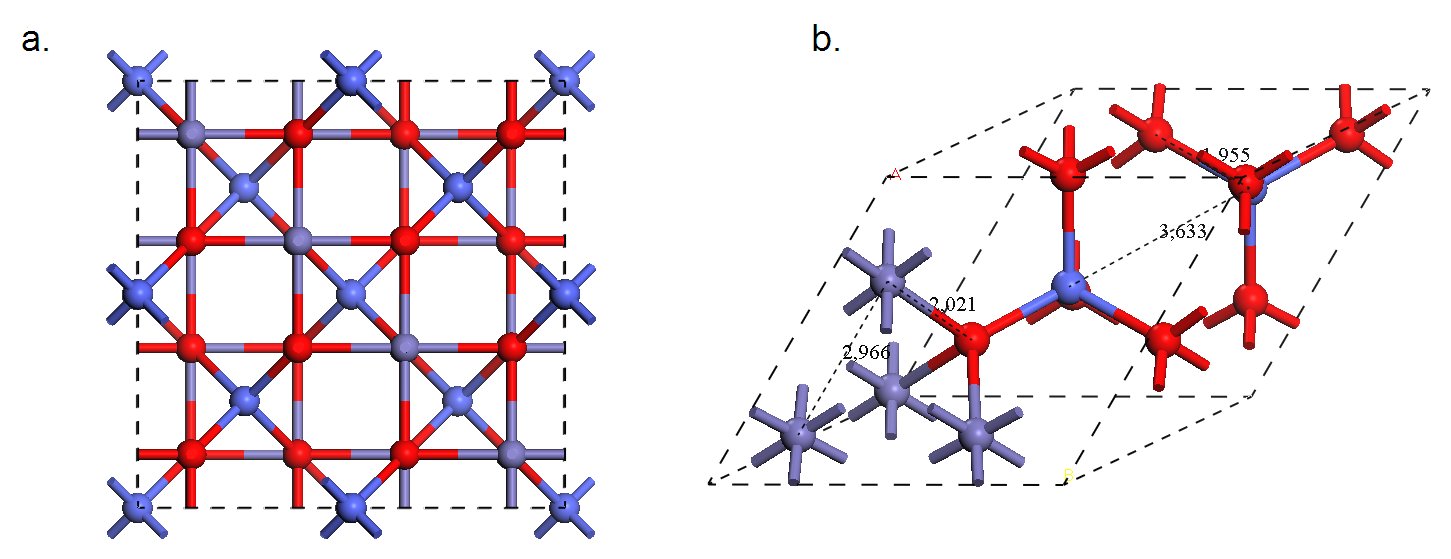}
\caption{The constructed unit cell of CoFe$_2$O$_4$ a) along with the primitive cell, b) used for the vibrational analysis.\label{fig4}}
\end{figure}

	Periodic Boundary Conditions (PBC) calculations of the phonon frequencies for were performed with CASTEP code as implemented in Materials Studio 5.5 \cite{journal-15},\cite{journal-16} and Generalized Gradient Approximation (GGA) represented by PBE functional \cite{journal-17,journal-18}. The computations were performed with the Troullier-Martins type Norm-Conserving (NC) pseudopotentials developed by Bennett  and  Rappe \cite{journal-19} with the Plane-Wave kinetic energy cut-off of 750 eV. First, the unit cell consisted of 56 atoms, within the well known cubic spinel structure (space group = $Fd-3m$; $a_0$ = 8.385~\AA), was constructed as presented in Fig. \ref{fig4}a. In the next step the unit cell was reduced into the Co$_2$Fe$_4$O$_8$ primitive cell of face-centered cubic. The resulted lattice parameters were fixed during the optimization and the delocalized internals were relaxed with the 5.0$\times$10$^{-7}$ eV/atom; 5.0$\times$10$^{-3}$ eV/\AA~and 5.0$\times$10$^{-3}$~\AA~convergence criteria for the energy; Hellman-Feyman forces and atomic displacements, respectively. The electronic energy was computed within the precise (54$\times$54$\times$54) FFT grid using the 5$\times$5$\times$5 Monkhorst-Pack grid for the Brillouin-zone sampling. The SCF tolerance of 5.0$\times$10$^{-10}$~eV/atom was applied along with the electron smearing of 0.1 eV. The resulted equilibrium structure was presented in Fig.~\ref{fig4}b. Finally, the computations were also repeated under the same conditions with the full size unit cell presented in Fig.~\ref{fig4}a.	

Since the studied structure is not an insulator it was impossible to simulate the IR activities via the linear-response approach. Also applying the Hubbard correction, what might be suggested for the studied system, is still impossible in phonon calculations using the available solid state DFT software. Hence, the numerical finite-displacement approach, as implemented in CASTEP, was applied to calculate the phonon modes at the $\Gamma$ point as seen by vibrational optical spectroscopy. For the calculated frequencies the Acoustic Sum Rule (ASR) has been applied. No imaginary frequencies were presented in the calculated results. It should be noticed that we have also tested the ultrasoft pseudopotentials along with pure LDA approach however no improvement was found in this case.

\subsection{Vibrational Analysis}
According to group theory, the irreducible representations for the studied systems are as follows: 

$$\Gamma_{red} = A_{1g}+E{_g} +T_{1g}+3T_{2g}+2A_{2u}+2E_u +5T_{1u}+2T_{2u}$$

Within the expected normal modes, only the T${_1u}$ type vibrations are infrared-active, while A$_{1g}$, E$_g$ and T$_{2g}$ symmetry modes are allowed in Raman spectroscopy. All the IR-active vibrations are triply degenerated. T$_{1g}$, A$_{2u}$, E$_u$ and T$_{2u}$ symmetry vibrations are the silent ones. None of the modes may be observed both by infrared and Raman spectroscopy. As the complete unit cell consists of 56 atoms, the number of the modes allowed to be observed with optical vibrational spectroscopy equals 165 (3N-3). By using the primitive cell approach we have reduced the computational problem into 39 modes which undergo further degeneracy upon expanding the system. However, the computations revealed that the calculations within the full-unit cell dimension resulted mainly in further splitting of the silent modes, with no significant influence on the optically active vibrations.

\begin{figure}
\includegraphics[width=1.0\textwidth]{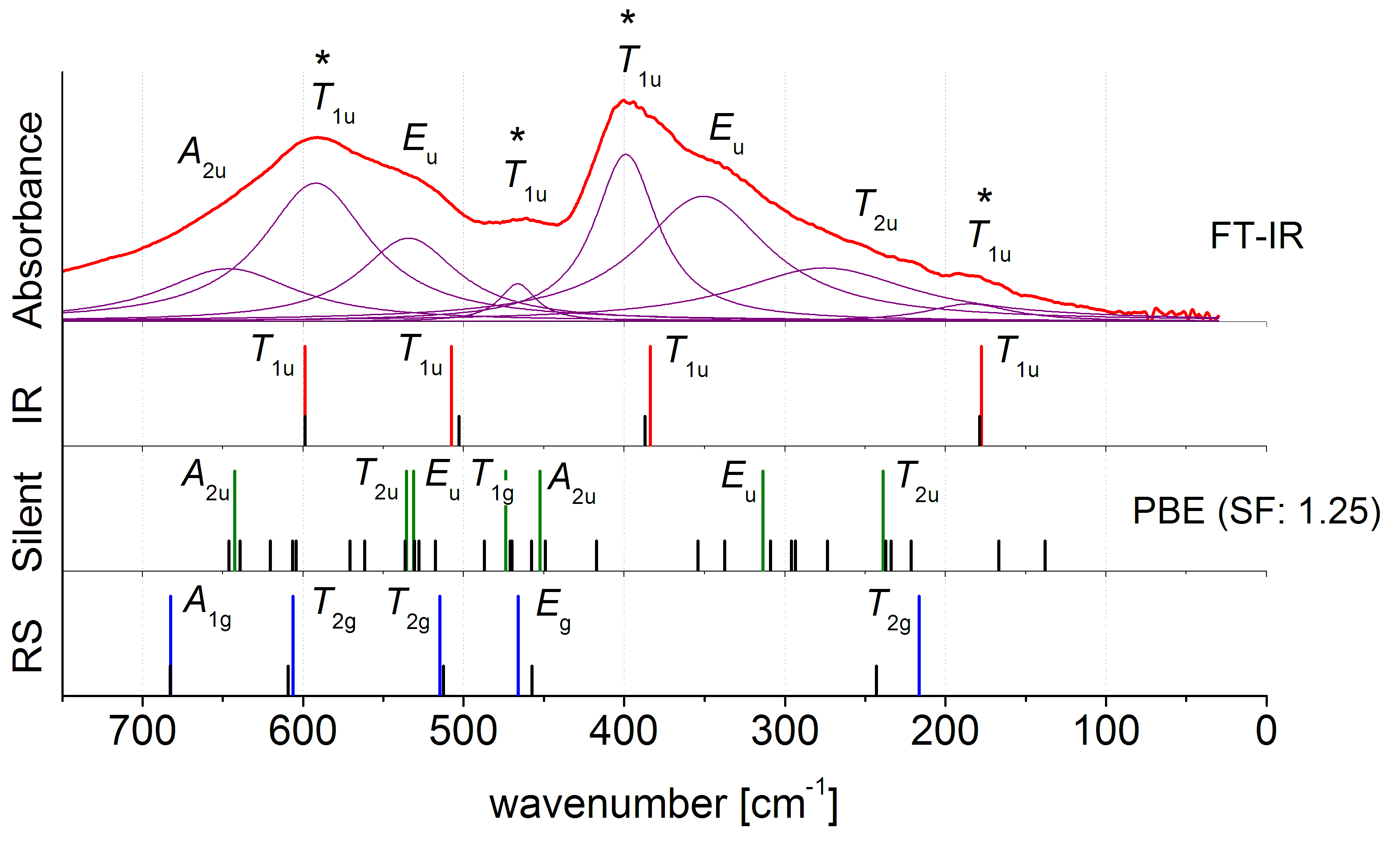}
\caption{Experimental FT-IR spectrum of CoFe$_2$O$_4$ recorded at room temperature along with the theoretical frequencies of the IR- and RS-active and the silent modes. The higher sticks correspond to the primitive-cell model, while the lower ones comes from the full unit cell approach. The main infrared bands are denoted with an asterisk. Note that the computed frequencies were multiplied by 1.25 factor.\label{fig5}}
\end{figure}

Fig. \ref{fig5} compares the experimental room temperature FT-IR spectrum of the CoFe$_2$O$_4$ sample, with the calculated frequencies of the IR and Raman active modes and with the symmetry forbidden vibrations (silent modes). The higher sticks correspond to the primitive-cell model, while the lower ones are due to the full-unit cell approach. The infrared active T$_{1u}$ type modes are visualized in Fig.~\ref{fig6}. The projection of all the calculated phonons has been given in the supplementary section.

\begin{figure}
\includegraphics[width=1.0\textwidth]{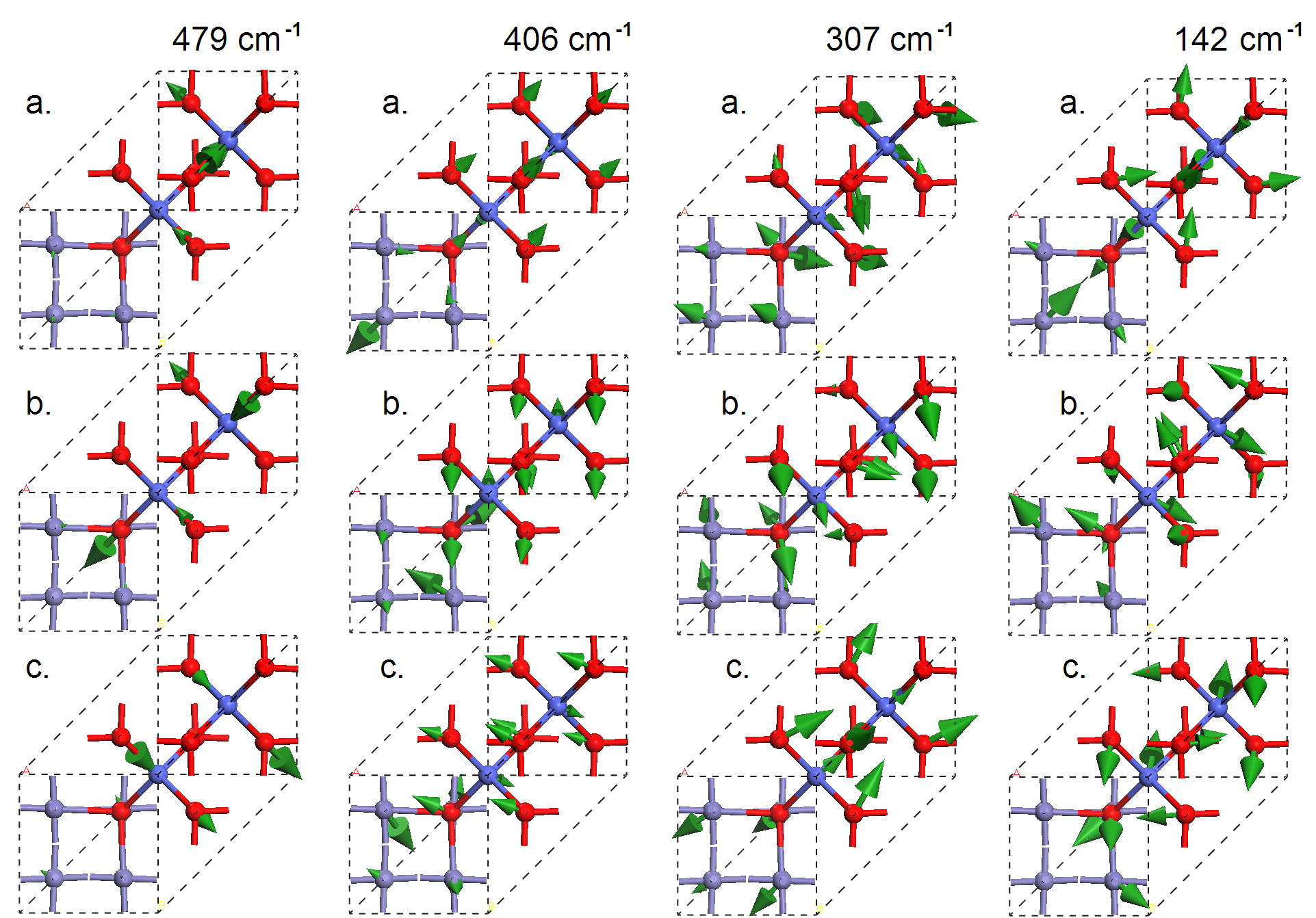}
\caption{The infrared-active, triply-degenerated (a,b,c) T$_{1u}$ type modes of CoFe$_2$O$_4$ primitive cell along with the corresponding frequencies calculated with PBE/NC/750~eV.\label{fig6}}
\end{figure}

All the theoretical frequencies fall into the far-infrared regime. The theory predicts four, triply degenerated modes, observed at: 479, 406, 307 and 142 cm$^{-1}$. However, the experiment reveals more spectral features. Figure~\ref{fig5}. delivers the experimental spectrum deconvoluted with Lorentzian functions giving the following bands: 646, 592, 534, 466, 399, 351, 275 and 185 cm$^{-1}$. The calculated frequencies are strongly underestimated with respect to the experiment for about $\sim$25\%. This effect may be partially linked with the insufficiency of the pure DFT in the description of such strongly electron-correlated system. Although the obtained results suggest a need of introducing a better theoretical model, it allows for sufficient interpretation of the experimental features. Hence, we have empirically rescaled the calculated frequencies by the scaling factor of 1.25. The computations suggest that the T$_{1u}$ type modes may be assigned to the 592, 466, 399 and 185 cm$^{-1}$ bands. The 534 and 351 cm$^{-1}$ features seems to be the shoulders of the adjacent T$_{1u}$ bands. Although the experimental spectrum may be disturbed by a modulated background, we believe, that the remaining spectral features comes from the activation of the A$_{2u}$, E$_u$ and T$_{2u}$ type silent modes, induced by a distortion of the structure which lowers the symmetry. However, we cannot also exclude that the shoulder bands may be also related with a further resonance splitting of the main T$_{1u}$ modes due to some long-correlation effects.

\section{Conclusions}

FT-MIR and FT-FIR measurements were performed for the first time for CFO ceramic in the wide temperature range. In this paper it is shown that CFO is a structural stable compound in the temperatures from 8~K~to~300~K. The small and large peaks of the bands are visible without changes in a wide temperature range. Moreover, none new band with wavenumber higher than 1000 cm$^{-1}$ appears. It means that these types measurements did not discover any phase transitions for the CFO ceramic. It agrees with other data~\cite{journal-10}. Interpretation of Raman spectra in the temperature range (300~-~870)~K~\cite{journal-13} and for CFO particles~\cite{journal-14}, and single crystal~\cite{journal-8} can be used in this case.  The obtained results of the measurements of the lead-free CoFe$_2$O$_4$ ceramic suggest the wide possibility of practical application of this material.

\subsubsection*{Acknowledgments}

The infrared (MIR and FIR) researches were carried out with the equipment purchased thanks to the financial support of the European Regional Development Fund in the framework of the Polish Innovation Economy Operational Program (contract no. POIG.02.01.00-12-023/08). The calculations were done at the Academic Computer Centre CYFRONET AGH, Cracow, Poland (Grants ID: MNiSW/IBM\_BC\_HS21/UJ/032/2009 – Mars Supercomputer) using Materials Studio 5.5 within Accelrys \textregistered polish national license.

\subsubsection*{Supplementary Information Available}
Supporting information about the vibrational analysis of CoFe$_2$O$_4$ may be found in the online version of this article, presenting the visualisation of all the normal vibrations. This information is available free of charge via the Internet at http://....

\end{document}